\documentclass[aps,twocolumn,prl,superscriptaddress]{revtex4-1}

\usepackage{amsfonts,amssymb,bm}
\usepackage{amsmath}
\usepackage{graphicx}
\usepackage[english]{babel}
\usepackage{graphicx, lettrine}
\usepackage{natbib}
\usepackage{mathrsfs}
\usepackage{color}
\usepackage{subcaption}
        
\begin{document}
\newcommand{\red}[1] {\textcolor{red}{#1}}
\newcommand{\ket}[1]{\left| #1 \right\rangle}
\newcommand{\bra}[1]{\left\langle #1 \right|}
\newcommand{\braket}[2]{\left\langle #1 | #2 \right\rangle}
\newcommand{\braopket}[3]{\bra{#1}#2\ket{#3}}
\newcommand{\proj}[1]{| #1\rangle\!\langle #1 |}
\newcommand{\expect}[1]{\left\langle#1\right\rangle}
\newcommand{\Tr}{\mathrm{Tr}}
\def\Id{1\!\mathrm{l}}
\newcommand{\cM}{\mathcal{M}}
\newcommand{\cR}{\mathcal{R}}
\newcommand{\cE}{\mathcal{E}}
\newcommand{\cL}{\mathcal{L}}
\newcommand{\cl}{l}
\newcommand{\cH}{\mathcal{H}}
\newcommand{\cU}{\mathcal{U}}
\newcommand{\cP}{\mathcal{P}}
\newcommand{\reals}{\mathbb{R}}
\newcommand{\grad}{\nabla}
\newcommand{\rhohat}{\hat{\rho}}
\newcommand{\rhoMLE}{\rhohat_\mathrm{MLE}}
\newcommand{\rhotomo}{\rhohat_\mathrm{tomo}}
\newcommand{\diff}{\mathrm{d}\!}
\newcommand{\pdiff}[2]{\frac{\partial #1}{\partial #2}}
\newcommand{\todo}[1]{\color{red}#1}
\def\FCW{1.0\columnwidth}
\def\HCW{0.55\columnwidth}
\def\TPW{0.33\textwidth}
\def\tred#1{\textcolor{red}{#1}}
\def\cred#1{\textcolor{red}{(#1)}}

\newcommand\blfootnote[1]{%
	\begingroup
	\renewcommand\thefootnote{}\footnote{#1}%
	\addtocounter{footnote}{-1}%
	\endgroup
}

\title{A SOI Integrated Quantum Random Number Generator Based on Phase fluctuations from a Laser Diode}

\author{Francesco Raffaelli}
\address{Quantum Engineering Technology Labs, Department of Physics, Tyndall Avenue, University of Bristol, Bristol, United Kingdom, BS8 1TH}
\author{Philip Sibson}
\address{Quantum Engineering Technology Labs, Department of Physics, Tyndall Avenue, University of Bristol, Bristol, United Kingdom, BS8 1TH}
\author{Jake E. Kennard}
\address{Quantum Engineering Technology Labs, Department of Physics, Tyndall Avenue, University of Bristol, Bristol, United Kingdom, BS8 1TH}
\author{Dylan H. Mahler}
\affiliation{Quantum Engineering Technology Labs, Department of Physics, Tyndall Avenue, University of Bristol, Bristol, United Kingdom, BS8 1TH}
\address{Now at: Xanadu, 372 Richmond St W, Toronto ON, M5V 1X6, Canada}

\author{Mark G. Thompson}
\address{Quantum Engineering Technology Labs, Department of Physics, Tyndall Avenue, University of Bristol, Bristol, United Kingdom, BS8 1TH}
\author{Jonathan C. F. Matthews}
\address{Quantum Engineering Technology Labs, Department of Physics, Tyndall Avenue, University of Bristol, Bristol, United Kingdom, BS8 1TH}

\begin{abstract}
Random numbers are a fundamental resource in science and technology. Among the different approaches to generating them, random numbers created by exploiting the laws of quantum mechanics have proven to be reliable and can be produced at enough rates for their practical use. While these demonstrations have shown very good performance, most of the implementations using free-space and fibre optics suffer from limitations due to their size, which strongly limits their practical use. Here we report a quantum random number generator based on phase fluctuations from a diode laser, where the other required optical components are integrated on a mm-scale monolithic silicon-on-insulator chip. Our device operates with generation rate in the Gbps regime and the output random numbers pass the NIST statistical tests. Considering the device's size, its simple, robust and low power operation, and the rapid industrial uptake of silicon photonics, we foresee the widespread integration of the reported design in more complex systems.    
\end{abstract}


\maketitle

\section{Introduction}
In the last few decades \cite{Herrero-Collantes2017}, quantum random number generators (QRNG) have been demonstrated with different optical systems such as single photons \cite{qrng,jennewein2000}, optical vacuum states \cite{Gabriel2010,high_bitrate_qrng,avesani2017,Xu2017} and phase fluctuations from a laser diode \cite{Qi2010,ultrafast_qrng,68gbps_qrng,Jofre2011,Abellan2014}.  
Among all these schemes, the QRNGs based on phase fluctuations from a laser diode achieved the highest generation rates, up 68 Gbps \cite{68gbps_qrng}. 
 On one hand, this fast development shows the relevance of QRNG to modern science and technology. On the other hand, it must be noticed how all the aforementioned works were performed either with bulk or fibre components, leading to strong practical limitations. First, given the size of the components used, these devices are difficult to integrate into more complex systems, such as a Quantum Key Distribution (QKD) receiver or a CPU. Second, the cost of each component is considerable, severely reducing scalability. Finally, bulk fibre optics devices often suffer from instability issues, limiting their use in real-world scenarios. 
 
In order to build more practical devices, the community started looking into ways to reduce the size of QRNGs. Sanguinetti et al. \cite{phone_qrng} demonstrated a QRNG taking advantage of a smartphone camera. Abellan and co-workers \cite{InP_qrng} demonstrated a random number generator fully integrated into a Indium Phosphide (InP) microchip and Raffaelli et al. \cite{Raffaelli2018} demonstrated a Silicon-on-Insulator (SOI) QRNG, using the scheme proposed by Gabriel and colleagues \cite{Gabriel2010}. Interestingly \cite{phone_qrng} and \cite{Raffaelli2018} were performed in the same platforms used for the first prototypes of integrated  QKD devices \cite{sibson2016,sibson2017,Ding2017}, supporting potential integration in the future. Recently, Haylock et. al \cite{Haylock2018} demonstrated multiplexing of the scheme reported in \cite{Gabriel2010}, in the Lithium Niobate (LN) platform --- this multiplexed approach allows enhanced generation rates, but the relatively large footprint of LN can limit the possibility of integrating LN chips into more complex systems. In this sense, the InP based QRNG \cite{InP_qrng} provides a more attractive solution, but it has the extra requirement of RF modulation to control the laser, which in turn limits the generation rate \cite{Shi-Hai2017}. The experiment by Raffaelli et al., based on homodyne measurements, does not present this complication. However, it does require optical powers one order of magnitude higher than those used in laser diode phase fluctuations implementations \citep{ultrafast_qrng,68gbps_qrng}. This can be problematic since high optical powers in integrated photonics introduces negative side effects such as temperature variations and optical cross talk, potentially detrimental if integrated in systems with a high density of components.  Here, in order to overcome these limitations, we combined the advantages of \cite{ultrafast_qrng,68gbps_qrng} with the ultra-small footprint and versatility of the silicon-on-insulator platform,  reporting the demonstration of a SOI integrated QRNG based on phase fluctuations from a diode laser. 

\section{Results}
\begin{figure*}[t!]
	\centering
	\includegraphics[width=.8\textwidth]{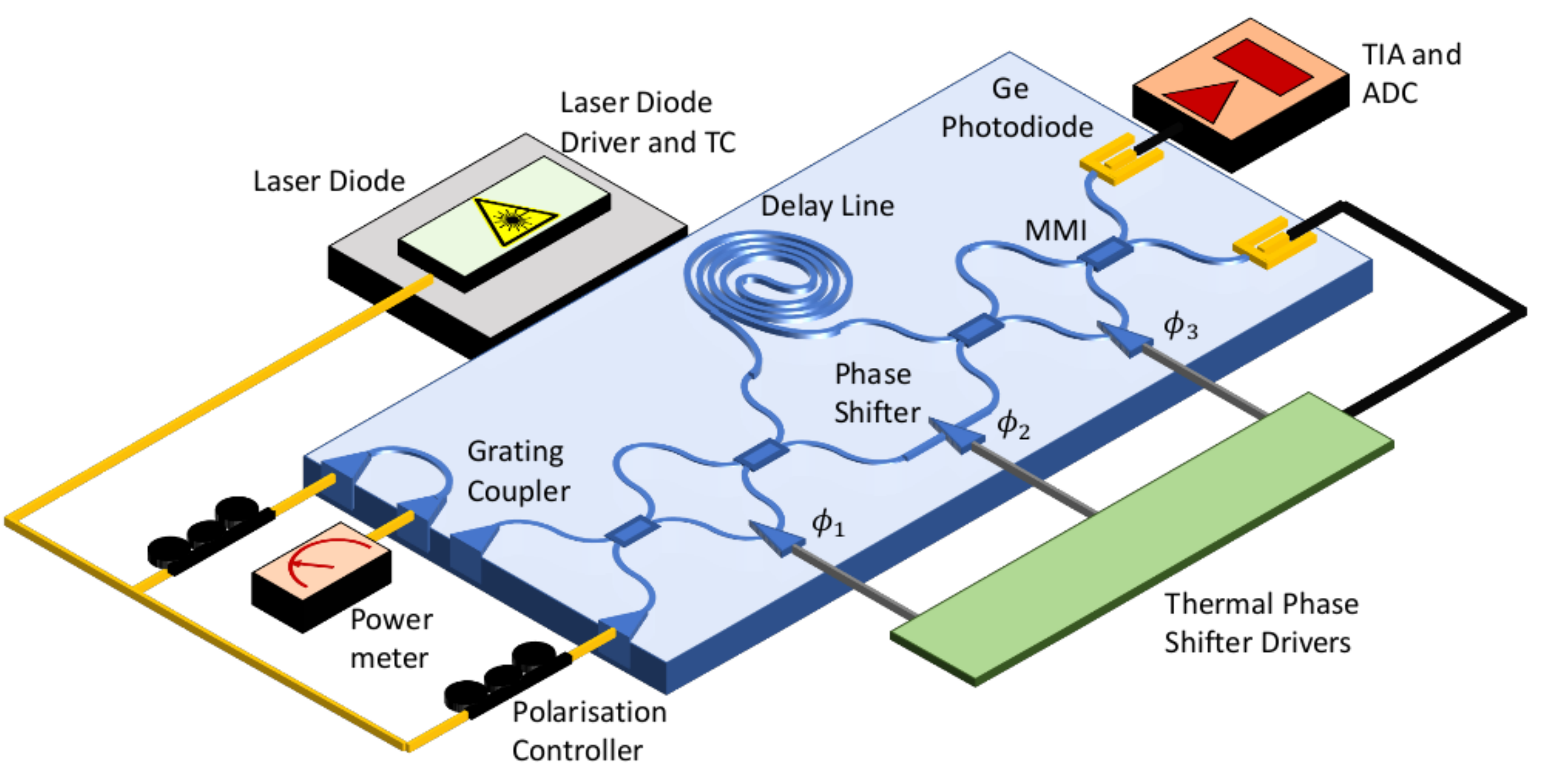} 
	\captionsetup{justification=raggedright, singlelinecheck=false}
	\caption{Setup of the experiment. a) A diode laser, controlled by a laser diode driver and temperature stabilised by a temperature controller, is used to operate just above threshold. A minimal part of the light is sent to a polarisation controller and coupled into a test waveguide, to monitor the coupling losses. The rest of light is sent to a polarisation controller and then coupled into the cascade of Mach-Zehnder interferometers. While the first and last MZIs serve as tunable beam-splitter, the central, unbalanced MZI converts the phase-fluctuations into intensity fluctuations. Two photodiodes are place at the output of the cascade. One of them is used as a monitor, and it allows to calibrate the phase of the interferometers. This is achieved through heater drivers, which control the phase of the MZIs by applying voltage to the integrated phase-shifters. The second photodiode is connected to a transimpedance amplifier, converting the light intensity fluctuations into voltage fluctuations. These fluctuations are digitalised by a oscilloscope to  generate random bits.}
	\label{setup_phase_fluctuations}
\end{figure*}
The technique used here lies on the intrinsic properties of laser emission. For any laser, the emitted light is given by a contribution from the stimulated emission and a contribution from the spontaneous emission \cite{Petermann1988,linewidth_semiconductor_laser,occupation_fluctuation_noise}.  The spontaneous emission, characterised by random phase fluctuations, can be efficiently exploited to generate true random numbers \cite{ultrafast_qrng,68gbps_qrng}. 
The electromagnetic field of the emitted light from a laser diode can be expressed as
\begin{equation}
E(t)=Ee^{-i({\omega}t+\theta(t))},
\end{equation}  
where $\omega$ is the angular frequency of the electromagnetic field and $\theta(t)$ is a random phase due to the contribution of the spontaneous emission to the emitted light. In order to take advantage of the random phase-fluctuations of the electromagnetic field, the light is injected into an unbalanced Mach-Zehnder interferometer (MZI). After removing the DC components,  the intensity at the photodiodes will be given by
\begin{equation}\label{fringes_equation}
I(t) \propto P\sin(\Delta{\theta(t)}) \sim P\Delta{\theta(t)},
\end{equation} 
where the first equation holds when the phase delay due to the different length between the two arms of the MZI is $2m\pi+pi/2$, and the second relation is valid for small values of $\Delta{\theta(t)}$. The light intensity at the photodiodes is converted into voltage signal by a transimpedance amplifier and the variance of the voltage measured by the oscilloscope is
\begin{equation}\label{voltage_variance}
\sigma^2 \equiv \langle{\Delta{V(t)}^2}\rangle \propto AP^2\langle{\Delta{\theta(t)}^2}\rangle + F,
\end{equation}  
where $\sigma^2 \equiv \langle{\Delta{\theta(t)}^2}\rangle$ is the variance of the phase noise, A is a conversion constant between the optical power and voltage variance, which takes into account the responsivity of the photodiodes and gain of the amplifying electronics, and $F$ is the variance of the background electronic noise. It can be shown \cite{linewidth_semiconductor_laser,occupation_fluctuation_noise} that the variance of the random phase of a diode laser is the sum of an intrinsic quantum phase noise $Q$ and a classical phase noise $C$ and it can be expressed as 
\begin{equation}
\langle{\Delta{\theta(t)}^2}\rangle = (\frac{Q}{P}+C).
\end{equation}
As a consequence, the variance of the voltage becomes
\begin{equation}
\sigma^2 = ACP^2+AQP+F,
\label{voltage_variance_phasefluctuation}
\end{equation} 
and the parameters AC, AQ and F, dependent on the specific laser and measurement setup, can be determined through a polynomial fit. As mentioned above, the phase noise intrinsic to the spontaneous emission is expressed by the parameter AQ in Eq.~\ref{voltage_variance_phasefluctuation}. Therefore, we define the Quantum-to-Classical Noise Ratio (QCNR) as 
\begin{equation}\label{QCNR} 
QCNR = \frac{AQP}{ACP^2+F}.
\end{equation}
\subsection{Description of the experimental setup}

In Fig. \ref{setup_phase_fluctuations} we report a scheme of our experiment. A Mitsubishi FU-68SDF-8 DFB laser diode, driven by a Thorlabs CLD1015 module, was used as a light source. The light was sent to a polarisation controller to optimise the coupling of the optical beam on chip. The vertical coupling was achieved by using a 8-channel V-groove array (VGA), coupled into the single mode waveguides with grating couplers. Part of the light, during the characterisation process, was sent through a test waveguide to a power meter for monitoring the coupling losses into the chip.
The SOI chip used for this experiment was designed using iSiPP25G technology and manifactured by IMEC \cite{iSiPP25G}. Our QRNG was characterised by  a  cascade of three Mach-Zehnder interferometers.
 Indeed, Eq. \ref{fringes_equation} is based on the assumption of perfectly balanced beam-splitter and lossless optical channel. When working in bulk and fibre optics these assumptions are satisfied to a very high degree. However, fabrication errors in integrated SOI devices can drastically alter the reflectivity. For example, the integrated  multi-mode interferometers (MMIs) used in our experiment have show around  0.5 dB excess loss. Moreover, the linear losses in our single mode waveguides are estimated to be 2-3 dB/cm, not negligible in a long delay line. This implies that the physical device must be designed in such a way to take into account the imperfections in the integrated device. Therefore, the input and output MZIs, controlled by $\phi_1$ and $\phi_3$, were balanced MZIs, where the length of the two arms was equal, and the relative phase between the two arms could be tuned by taking advantage of a thermal phase-shifter. These MZIs formed tunable reflectivity beam-splitters. The central MZI, controlled by $\phi_2$, was instead unbalanced, with a time delay $T_d \sim $ 540  ps (corresponding to $\sim$ 4cm length) between the two arms, which allowed mapping the fluctuations in the phase of the electromagnetic field into intensity fluctuations. This central MZI presented a phase-shifter used to configure the system to optimise the intensity fluctuations. 
 
While most of the waveguides were standard strip single mode waveguides with a $220nm\times 450nm$ cross section, the delay line was characterised by a broader rib waveguide to limit the losses.  The MZIs cascade was designed with two Ge photodiodes at the outputs. The first photodiode was used to monitor the optical power, and it could be used to vary the phase of the integrated phase-shifters, through heater drivers controlled via computer.  The photo-current from the second photodiode was converted into a voltage signal and amplified by a custom made high-speed, low-noise transimpedance amplifier (TIA). The voltage signal was detected and digitalised by a fast GHz bandwidth oscilloscope (DSOV134A Agilent Keysight Technology) and the data were further analysed and post-processed by a desktop computer, to extract random bits. In practise, the grating couplers, MZI and photodiodes  occupied an area  $<$ 1$mm^2$, integrated on a SOI chip with a footprint of 2.5mm$\times$2.5mm. The chip was embedded and wirebonded to a 4cm$\times$8cm electronic printed circuit board, containing the TIA and the voltage supply for the photodiodes. This system was enclosed inside a Faraday cage, to reduce the RF environmental noise.

\subsection{Determination of the Quantum-to-Classical Noise Ratio QCNR}
\begin{figure*}[htbp!]
	\centering
\includegraphics[width=1\textwidth]{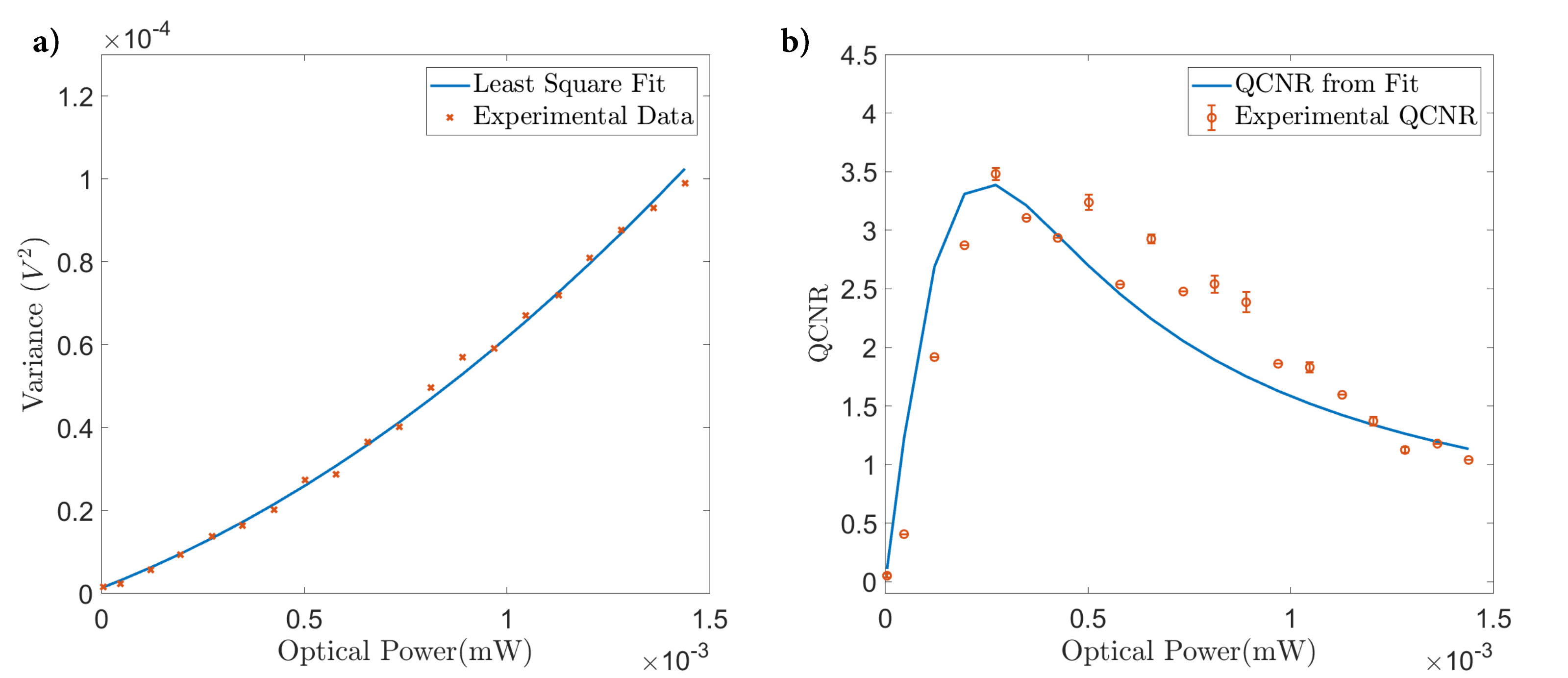} 
	\captionsetup{justification=raggedright, singlelinecheck=false}
	\caption{Phase-fluctuations variance and QCNR as a function of the optical power. a) Quadratic fit of the noise variance. b) In blue the QCNR extracted by the quadratic fit. In orange, the experimentally determined QCNR.}
	\label{variance_QCNR_vs_power}
\end{figure*}

To estimate the QCNR, for each value of the input current, the maximum phase noise variance had been extracted after scanning the phase $\phi_2$ over $2\pi$ (see Supplementary Material for more details). Then, the variance was plotted as a function of the optical power, shown in Fig. \ref{variance_QCNR_vs_power}(a). After this, the parameters AC, AQ and F were determined by using a quadratic least square algorithm,  
and consequently it was possible to extract the QCNR, reported in Fig. \ref{variance_QCNR_vs_power}(b) (blue continuous line).
The parameter of the fit are reported in Table \ref{fit_parameters}.

\begin{table}[htbp!]
\centering 
\setlength{\tabcolsep}{8pt}
\captionsetup{justification=raggedright,singlelinecheck=false}	
\caption{The statistical parameters for the fit of the voltage noise variance as a function of the optical power.}
\begin{tabular}{lc} 
\hline
 Parameter  &  Value     \\ [0.5ex]
\hline 
AC($V/W^2$)                 &  22.519     \\
AQ($V/W$)           &  0.03784  \\
F ($V$)           &  1.3732$\times 10^{-6}$   \\ [0.5ex]
\hline

R-Squared                       &  0.998    \\ 
\hline
\end{tabular}

\label{fit_parameters} 
\end{table}
A second experimental approach, suggested in Ref. \cite{ultrafast_qrng}, was used to verify the QCNR. This second approach has the main advantage that the QCNR can be estimated experimentally for a single given value of the power. It relies on the fact that the phase-fluctuations are due to a quantum contribution, that we called Q and a classical contribution that we called C. The quantum contribution will be dominant when the laser is operated just above threshold, while the classical contribution will be dominant when the laser diode is operated with a high input currents.  Hence, in order to estimate the QCNR experimentally, first of all we
measured the variance $\sigma^2$ for a given power $P_0$. This was then compared to the variance $\sigma^2_{att}$ measured when the laser was operated in the classical regime (at maximum power) and attenuated with a variable optical attenuator (VOA) to obtain the same optical power $P_0$.  

Finally, we calculated the ratio $QCNR_{exp}=\frac{\sigma^2-\sigma^2_{att}}{\sigma^2_{att}}$ which gave the ratio between the pure quantum contribution and the contribution due to the classical noise. The experimentally measured QCNR is shown in Fig.~ \ref{variance_QCNR_vs_power}(b) (red circles). Here we observe a good agreement between the QCNR obtained with the two methods. The discrepancy for some points with the fit is due to the fact that, given the high input losses of the VOA, we had to manually remove the VOA to obtain $\sigma^2$. This operation slightly affected the polarisation and thus the coupled optical power into the chip. Here it is worth noting that the optical power, reported in the x-axis in Fig. \ref{variance_QCNR_vs_power}, is the optical power before coupling into the chip. This is one order of magnitude lower than the optical power used in \cite{Raffaelli2018}. This demonstration at lower optical powers allows mitigation against potentially negative effects such as self phase modulation and optical cross talk, making this scheme more suitable for integration in more complex systems.

\subsection{Estimation of the min-entropy $H_\infty$}
Similarly to \cite{post_processing_qrng}, in the order to estimate the maximum extractable randomness, the min-entropy of the digitalised voltage signal can be calculated.
\begin{equation}
\mathrm{H}_{\infty}=-\mathrm{log_2}( \underset{x\in \{0,1\}^{n}}{\mathrm{max}}{\mathrm{Pr}[X=x]})
\label{min-entropy_phase-fluctuations}
\end{equation} 
is the min-entropy, where n is the number of bits used in the digitalisation of the voltage signal and $\mathrm{Pr}[X=x]$ is the probability of the voltage measurement x, falling in the X bin.

It can be shown that $\Delta{V(t)}$ has a gaussian distribution, being the linear combination of three different gaussian contributions. For this reason, the voltage variance due to the quantum phase-fluctuations can be obtained as
\begin{equation}\label{voltage_variance_quantum}
\sigma_q^2 = \frac{\sigma^2}{1+\frac{1}{QCNR}}.
\end{equation}
The following procedure was used to determine the min-entropy.

 For different values of the optical power, we swept the phase of the central interferometer from 0 to $\pi$, recording the fringes of the variance (see Supplementary Data for a figure of the fringes).
Then, for each optical power, we selected the maximum variance from the fringe and we plotted the maximum variance as a function of the optical power. The next step consisted in extracting the QCNR by using Eq. \ref{QCNR}. Once we had determined the QCNR, we calculated $\sigma_q^2$, by making use of Eq. \ref{voltage_variance_quantum}.
We then chose a voltage range in the oscilloscope to optimise the information contained in the measured signal (in our case $V(max,min)=\pm 5\sigma$) and divided the interval into $2^8$ bins, due to the resolution of our oscilloscope.
Finally, we integrated the signal over the bins and normalise the distribution to obtain $Pr[x]$ as in Eq.~\ref{min-entropy_phase-fluctuations} and calculated the min-entropy $H_{\infty}$. As shown in the previous paragraphs, we obtained QCNR $\sim$ 3.38. As a combination of these factors, we estimated the min-entropy to be $H_{\infty} \sim 5.6 ~ bits/sample$. This value of the entropy was used to implement a software version of the Toeplitz extractor \cite{post_processing_qrng}, that allowed us to obtain uniformily distributed random bits.

\subsection{Bandwidth and generation rate estimation}
\begin{figure}[htbp!]
	\includegraphics[width=\columnwidth]{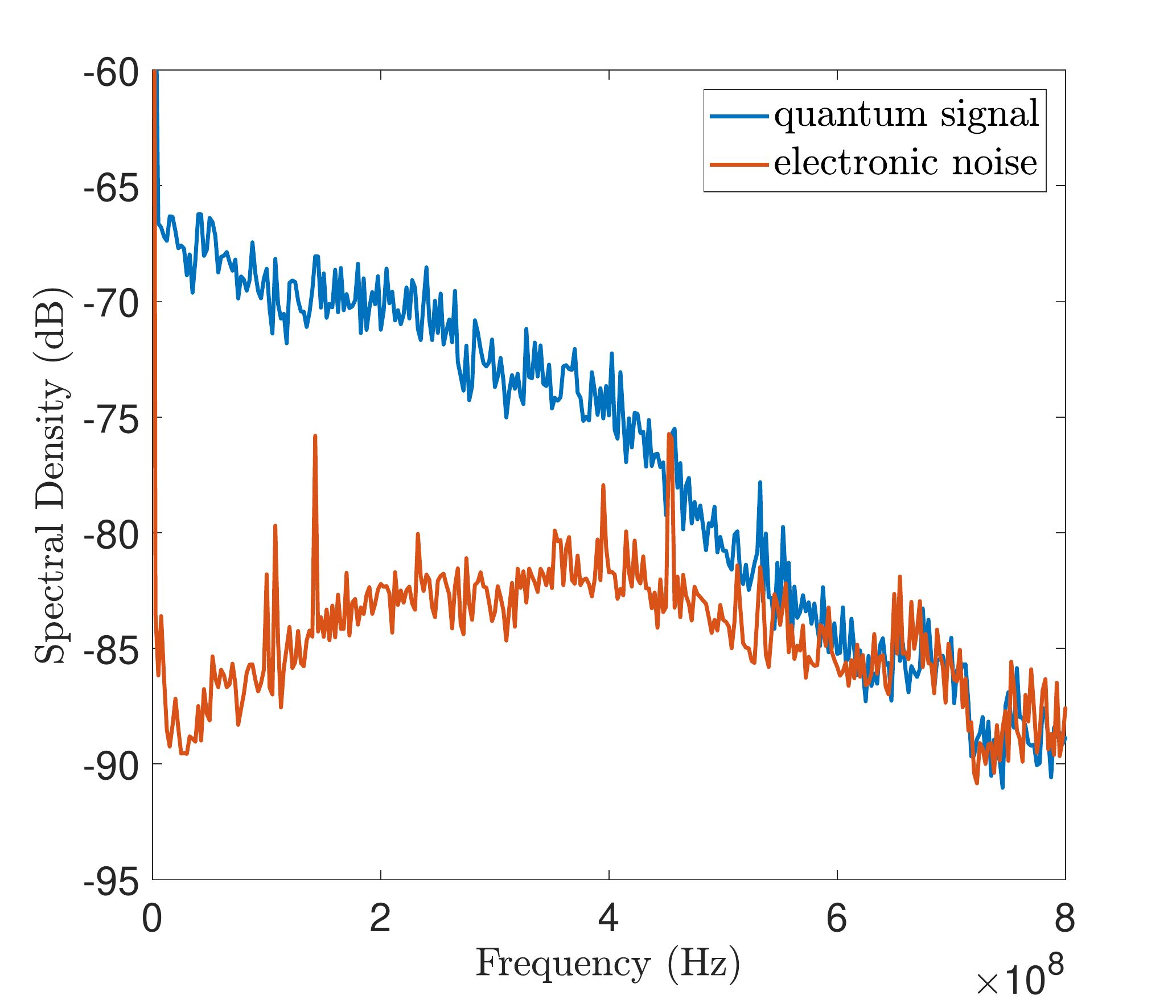} 
	\captionsetup{justification=raggedright, singlelinecheck=false}
	\caption{Spectral density for quantum signal and noise floor. Here the spectral density for the optical signal and for the electronic noise floor are reported. It can be observed that the quantum signal is above the electronic noise up to 500~MHz. The noise floor presents some peaks due to environmental noise, particularly  around 100~MHz. These are the FM radio signal, which however are below the quantum signal and therefore are not affecting the quality of the generated random numbers.}
	\label{spectrum_phase-fluctuations}
\end{figure}

In order to determine the optimal sampling rate of the device, the spectral density of our QRNG was measured in absence and presence of the optical signal. The result is reported in Fig. \ref{spectrum_phase-fluctuations}. Here we observe a bandwidth of approximately 500~MHz. From Fig. \ref{spectrum_phase-fluctuations} we can also see some peaks, mainly around 100~MHz, which are the radio environmental noise. However, the signal is well above the noise floor, so the environmental noise does not influence the generation of random bits. Taking into account a sampling rate of 500~Msamples, and that $H_{\infty} = 5.6~ bits/sample$, when sampling at 8~bits/sample, we estimated a potential randomness generation rate of nearly 2.8~Gbps. We note that the generation rate is more than one order of magnitude lower than in \citep{68gbps_qrng}. This is mainly due to the limited bandwidth of our low-noise TIA. A faster TIA combined with lower loss grating couplers and optimisation in the waveguides design would allow to increase the generation rate beyond 10 Gbps. 

\subsection{Autocorrelations and statistical tests}
\begin{figure*}[ht!]
	\centering
	\includegraphics[width=1\textwidth]{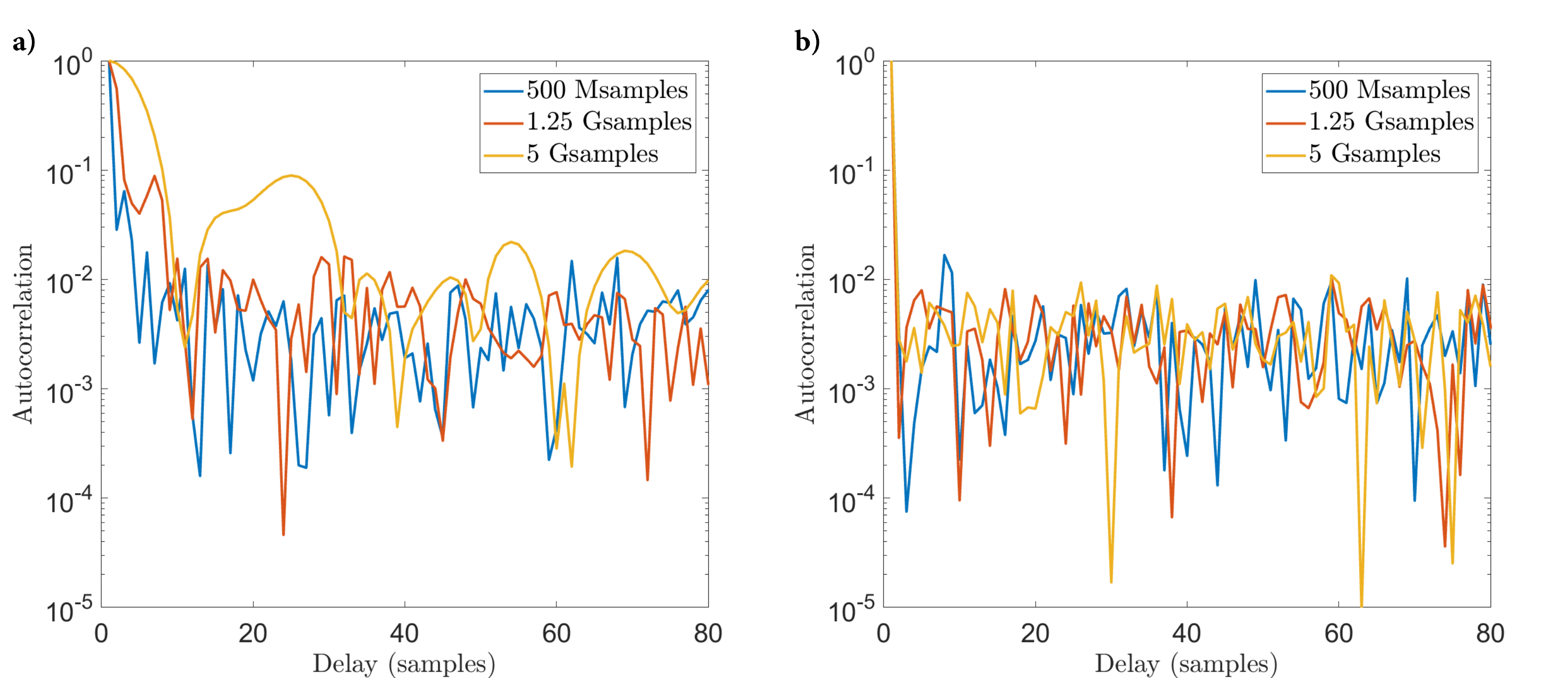} 
	\captionsetup{justification=raggedright, singlelinecheck=false}
	\caption{Autocorrelation of raw and hashed bits. a) Here we report the autocorrelations of the raw signal, obtained after digitizing the signal, for different sampling rates. We notice that for sampling speeds above the bandwidth the autocorrelations can be up to one order of magnitude bigger than the when sampling at 500 Msamples. b) The hashing reduces drastically the autocorrelations. No appreciable difference can be observed in this case between different sampling speeds.}
	\label{autocorrelation_phase-fluctuations}
\end{figure*}

A first estimation of the quality of the bit sequences is given by the autocorrelations of the samples. In fact, environmental RF noise, oscillations of the transimpedance amplifier and oversampling are the main cause of periodic oscillations in the signal that can result in correlated bit sequences. Moreover, since these factors can be in principle controlled classically by an adversary, it is important to study the autocorrelation of the signal to make sure about the unpredictability of the random bits. For this reason we measured the autocorrelation of the signal, acquired at different sampling rates. As expected, the optimal sampling rate appears to be 500 Msamples/s. This is because of the spectral density of the detector, where the optical signal is well above the electronic noise up to 500 MHz. On the other hand, oversampling at 5 Gsamples/s results in highly correlated sequences, as shown by the yellow line in Fig. \ref{autocorrelation_phase-fluctuations}(a). 

A second characterisation of the randomness was achieved by taking advantage of the statistical tests provided by the National Institute of Standards and Technology (NIST SP 800-22). As can be seen in Table \ref{nist_table}, the hashed bits passed all the statistical tests.

\begin{table}[htbp!]
\centering 
\setlength{\tabcolsep}{8pt}
\captionsetup{justification=raggedright,singlelinecheck=false}	
\caption{Here we report the results for the NIST (National Institute of Standards $\&$ Technology) statistical 
tests suite \cite{nist_web}. In order to pass the NIST SP800-22 the pass rate must be above 0.98 for each type of test  (column II) and the reported P-values, which refer to the uniformity test on the distributions plotted in Fig.~\ref{uniformity_test_phase-fluctuations}, must be above 0.01 (column III).}
\begin{tabular}{lcc} 
\hline
\textbf{NIST SP800-22} & & \\
\hline
 Test name  &  Pass Rate    &   P-value   \\ [0.5ex]
\hline
Frequency                  &  0.989   & 0.891 \\
Block Frequency            &  0.993   & 0.128 \\
Cumulative Sums            &  0.987   & 0.186 \\ 
Runs                       &  0.989   & 0.177 \\
Longest Run                &  0.992   & 0.768 \\
Rank                       &  0.995   & 0.360 \\ 
FFT                        &  0.984   & 0.465 \\
Non Overlapping Template   &  0.995   & 0.014 \\
Overlapping Template       &  0.986   & 0.155 \\
Universal                  &  0.985   & 0.800 \\
Approximate Entropy        &  0.984   & 0.573 \\
Random Excursions          &  0.993   & 0.115 \\
Random Excursions Variant  &  0.989   & 0.011 \\
Serial                     &  0.991   & 0.169 \\
Linear Complexity          &  0.993   & 0.768 \\ [1ex]
\hline
\end{tabular}
\label{nist_table} 
\end{table}

\begin{figure*}[htbp!]
	\centering
	\includegraphics[width=\textwidth]{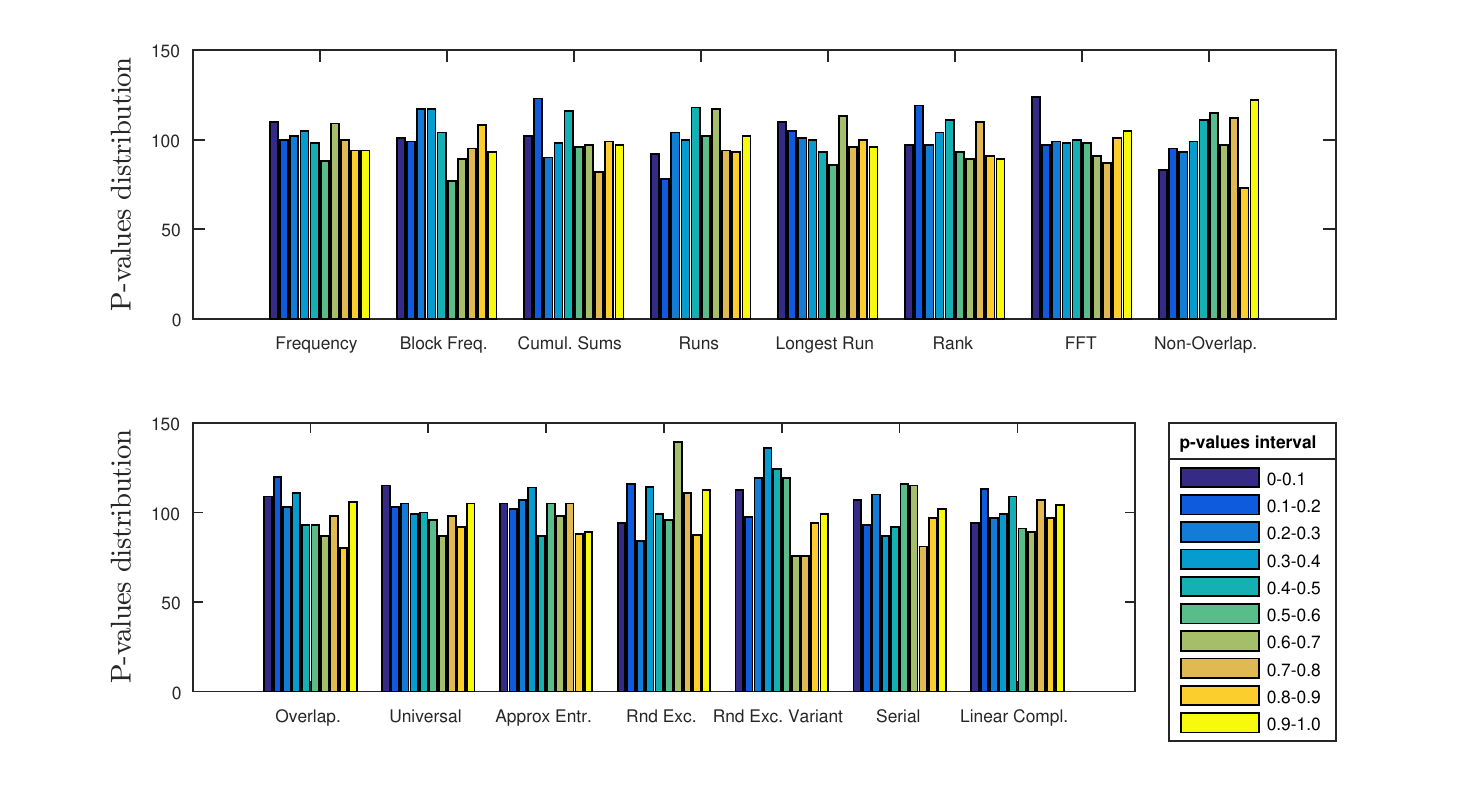} 
	\captionsetup{justification=raggedright, singlelinecheck=false}
    \caption{Uniformity test for the P-values. Under the assumption that the produced random   bits are truly random, the P-values must be uniformily distributed between 0 and 1. Here the NIST statistical test provides the frequencies of the P-values, by dividing the (0,1) interval  into 10 sub-intervals. We can observe that for each test the P-values are uniformly distributed.}	
	\label{uniformity_test_phase-fluctuations}
\end{figure*}

\subsection{Stability}
\begin{figure}[ht!]
	\includegraphics[width=\linewidth]{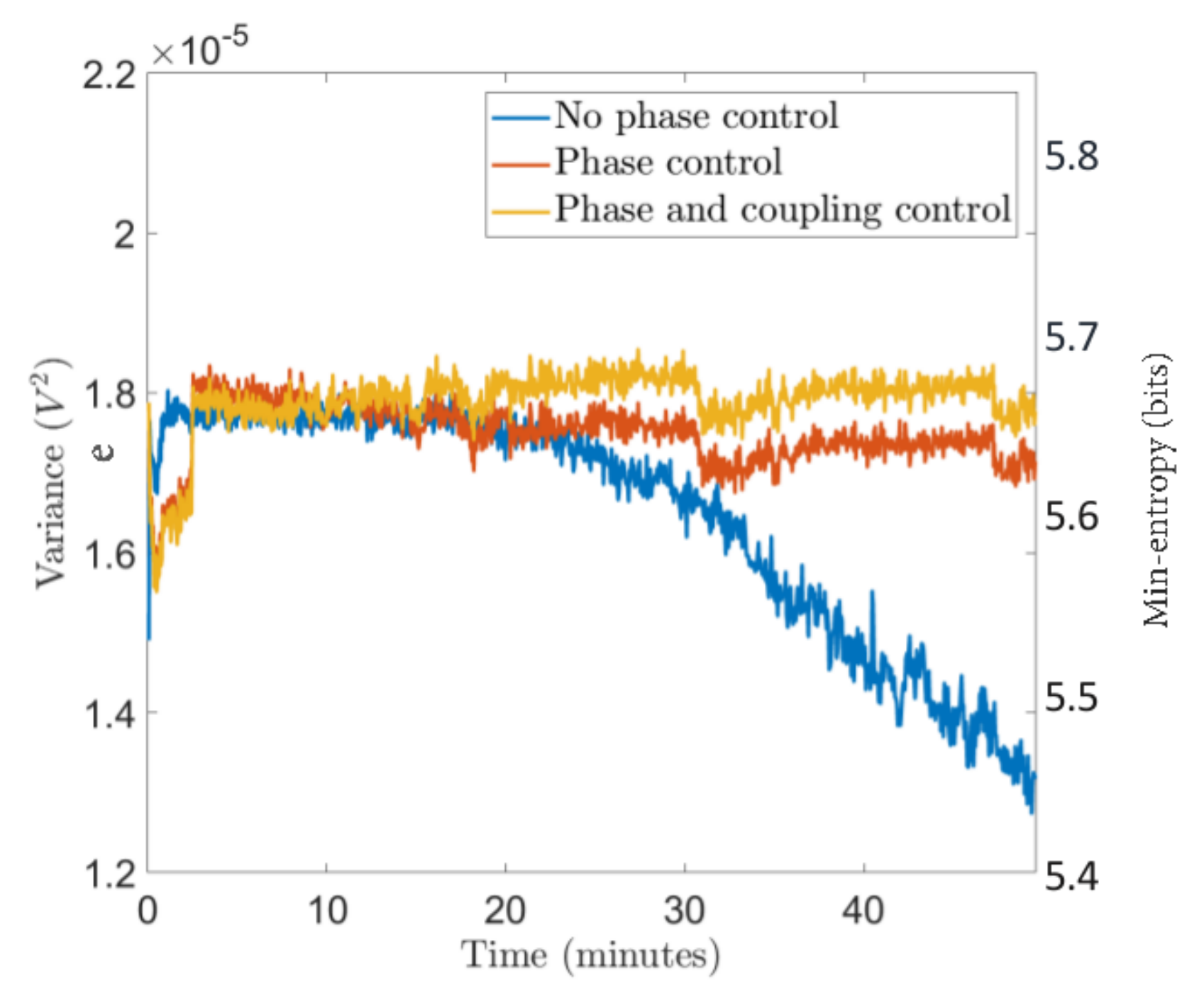} 
	\captionsetup{justification=raggedright, singlelinecheck=false}
	\caption{Signal variance measured over a time interval of one hour. In this picture the variance of the signal measured by the oscilloscope was measured over a time interval of 50 minutes. The blue line shows the behaviour of the variance without any control on the phase of the integrated MZIs. The red line shows the variance when the phase of the unbalanced MZI have been calibrated every 2 minutes. The yellow line has been obtained by normalising the variance plotted in red with the optical power coupled into the chip. It can be seen than, beside some variation due to the change in the fibre-chip coupling, a phase calibration every few minutes is sufficient to keep the system perfectly balanced.}
\label{stability_test}
\end{figure}
Among the main advantages of working with integrated photonics there are the potentially ultra-small footprint and monolithic nature of the devices. While compactness allows for parallelization of multiple components into a single microchip, the monolithic nature has the main advantage of strongly reducing many forms of instability. This is particularly useful when dealing with interferometry and unbalanced interferometers. For example, when working with optical fibres, small changes in temperature can affect the length of the fibre enough to destroy interference. In bulk optics instead, the stability is threatened by any environmental factor that generates oscillations in the optics. By reducing the size of the systems, and by integrating everything in a single chip, these issues are drastically reduced. This fact was particularly relevant in our experiment. In fact, 
as can be observed in the red line of Fig. \ref{stability_test}, the system was highly stable in a time range of almost one hour. This was obtained by simply calibrating the phase of the unbalanced interferometer every 2-3 minutes (red line). Moreover, even without any calibration in the phase of the interferometer, the variance of the signal (blue line), was stable over a time interval of several minutes. Here it is important to remark that  only the voltage in the unbalanced interferometer was scanned, while phase-shifters 1 and 3 remained untouched after a initial characterisation.  Furthermore, the yellow line was obtained normalising the variance taking into account the small variation in the optical power, due mainly to changes in polarisation in the light off-chip. In both the yellow and red lines, some steps in the variance can be observed. These steps are due to the fact that, while recalibrating the phase $\phi_2$, the point of maximum variance had slighly shifted. 
 In Fig. \ref{stability_test}, on the right vertical axis, we plotted the min-entropy obtained for that particular voltage variance. From this, it can be see that the system is very stable, allowing the entropy to be kept unchanged within the unity over a time interval of one hour, even without applying any phase control.

\section{Discussion}
In conclusion, we report the demonstration of a SOI integrated version of a QRNG based on phase fluctuations from a laser diode, where all the optical components and photodiodes are integrated onto a single monolithic microchip and fibre coupled to a laser diode.   We showed that high rates of random numbers can be achieved with sub mW optical powers, one order of magnitude lower the previous QRNG demonstrated in a SOI device \cite{Raffaelli2018}.  Compared to  \cite{68gbps_qrng,ultrafast_qrng}, our ultra-compact QRNG shows very high phase stability, strongly reducing the need for active stabilisation. While in \cite{InP_qrng}, the integrated laser source is an advantage in term of compactness and scalability, working with CW light, does not require RF modulation of laser diodes, simplifying the electronics design. 
A logical future direction of our demonstration is to use flip-chip bonding of VCSEL lasers to silicon photonics, as demonstrated recently \cite{Kaur15}. Moreover our QRNG is directly integrable into silicon devices, such as QKD systems demonstrated in \cite{sibson2017,Ding2017}. Alternatively, our device could be integrated into multimode SOI devices, as the one demonstrated by Wang et al. \cite{Wang2018}, to provide a true random seed for device-independent randomness expansion. Finally we expect our QRNG to find applications whenever a low impact, high rate source of random numbers will be required in Silicon-on-Insulator devices.

\section*{Funding Information}

This work was supported by ERC, PICQUE, BBOI, QUCHIP, the US Army Research Office (ARO) Grant No. W911NF-14-1-0133, the EPSRC Quantum Communications
Hub (EP/M013472/1) and the EPSRC programme grant (EP/L024020/1). MGT (EP/K033085/1) and JCFM (EP/M024385/1) acknowledge fellowship support from the EPSRC.

\section*{Acknowledgments}

The authors are grateful to M. Loutit, A. Murray and A. Crimp for technical support.



\end{document}